\documentstyle[proceedings,numreferences,mbf]{crckapb}

\def\openone{\leavevmode\hbox{\small1\kern-3.8pt\normalsize1}}%

\begin{opening}

\title{WHAT IS THE EVANS-VIGIER FIELD?}

\author{VALERI V. DVOEGLAZOV}

\institute{\it Escuela de F\'{\i}sica, Universidad Aut\'onoma de
Zacatecas\\
Apartado Postal C-580, Zacatecas 98068, Zac., M\'exico\\
E-mail:  valeri@ahobon.reduaz.mx\\
URL: http://ahobon.reduaz.mx/\~~valeri/valeri.htm}

\end{opening}

\begin{document}

\begin{abstract}
We explain connections of the Evans-Vigier model
with theories proposed previously. The Comay's criticism
is proved to be irrelevant.
\end{abstract}

\bigskip
\bigskip

The content of the present talk is the following:

\begin{itemize}

\item
Evans-Vigier definitions of the ${\bf B}^{(3)}$ field~\cite{EV1};

\item
Lorentz transformation properties of the ${\bf B}^{(3)}$ field and the
${\bf B}$-Cyclic Theorem\cite{DVO1,DVO2};

\item
Clarifications of the Ogievetski\u{\i}-Polubarinov, Hayashi and
Kalb-Ramond papers~\cite{Og,H,KR};

\item
Connections between various formulations of massive/massless
$J=1$  field;

\item
Conclusions of relativistic covariance and relevance of the Evans-Vigier
postulates.

\end{itemize}

In 1994-2000 I presented a set of papers~\cite{DVAB} devoted to
clarifications of the Weinberg (and Weinberg-like~\cite{Sankar,DVA1})
theories and the concept of Ogievetski\u{\i}-Polubarinov {\it notoph}.
In 1995-96 I received numerous e-mail communications from Dr. M.
Evans, who promoted a new concept of the longitudinal phaseless magnetic
field associated with plane waves, the ${\bf B}^{(3)}$ field (which
is later obtained the name of M. Evans and J.-P. Vigier). Reasons for
continuing the discussion during 2-3 years were: 1) the problem of
massless limits of all relativistic equations does indeed exist; 2)
the dynamical Maxwell equations have indeed additional solutions with
energy ${\cal E}=0$ (apart of those with ${\cal E} = \pm \vert
{\mbf\kappa}\vert$, see~\cite{Opp,Good,Gian,Ahl,Dvold};\footnote{If we
put energy to be equal to zero in the dynamical Maxwell equations
\begin{eqnarray}
{\mbf\nabla} \times [{\bf E} -i{\bf B}] + i (\partial/\partial t)
[{\bf E} - i{\bf B}] &=& 0\, ,\\
{\mbf\nabla} \times [{\bf E} +i{\bf B}] - i (\partial/\partial t)
[{\bf E} + i{\bf B}] &=& 0 \, .
\end{eqnarray}
we come to ${\mbf \nabla} \times {\bf E} = 0$ and ${\mbf
\nabla}\times {\bf B} = 0$, i.~e. to the conditions of longitudinality.
The method of deriving this conclusion has been given in~\cite{Dvap}.} 3)
the ${\bf B}^{(3)}$ concept met strong non-positive criticism (e.~g.,
ref~\cite{Lakh,Com,Hun}) and the situation became even more controversial
in the last years (partially, due to the Evans' illness).

What are misunderstandings of both the authors of the ${\bf B}^{(3)}$
model and
their critics? In {\it Enigmatic Photon} (1994), ref.~[1], the following
definitions of the longitudinal Evans-Vigier ${\bf B}^{(3)}$ field have
been given:\footnote{I apologize for not citing all numerous papers of
Evans {\it et al} and papers of their critics due to page restrictions on
the papers of this volume.}\\

{\it Definition 1.} [p.3,formula (4a)]
\begin{equation}
{\bf B}^{(1)} \times {\bf B}^{(2)} = iB^{(0)} {\bf B}^{(3)\,\star}\,,\quad
\mbox{et cyclic}.\label{bc}
\end{equation}

{\it Definition 2.} [p.6,formula (12)]
\begin{equation}
{\bf B}^{(3)}= {\bf B}^{(3)\,\star}= -{i\kappa^2\over B^{(0)}}
{\bf A}^{(1)} \times {\bf A}^{(2)}\,,
\end{equation}
and

{\it Definition 3.} [p.16,formula (41)]
\begin{equation}
{\bf B}^{(3)}= B^{(0)} \hat{\bf k}\,.
\end{equation}

The following notation was used: $\kappa$ is the wave number; $\phi=
\omega t - {\mbf \kappa}\cdot {\bf r}$ is the phase;
${\bf B}^{(1)}$
and ${\bf B}^{(2)}$ are usual transverse modes of the magnetic field;
${\bf A}^{(1)}$
and ${\bf A}^{(2)}$ are usual transverse modes of the vector potential.

The main experimental prediction of Evans~[1a,b] that the magnetization
induced during light-matter interaction (for instance, in the IFE)
\begin{equation}
{\cal M} = \alpha I^{1/2} +\beta I+ \gamma I^{3/2}\,\quad\mbox{where}\quad
I={1\over 2} \epsilon_0 c E_0^2, E_0 = c\vert{\bf B}_\pi\vert
\end{equation}
has {\bf not} been confirmed by the North Caroline group~\cite{NC}.
As one can see from Figure 4 of~\cite{NC} ``the behaviour of the
experimental curve does not match with Evans calculations".

Nevertheless, let us try to deepen understanding of the theoretical
content of the Evans-Vigier model. In their papers and books~[1] Evans and
Vigier used the following definition for the transverse antisymmetric
tensor field:
\begin{equation} \pmatrix{{\bf B}_\perp\cr {\bf
E}_\perp\cr} = \pmatrix{{B^{(0)}\over \sqrt{2}} \pmatrix{+i\cr 1\cr
0\cr}\cr {E^{(0)}\over \sqrt{2}} \pmatrix{1\cr -i\cr 0\cr}\cr} e^{+i\phi} +
\pmatrix{{B^{(0)}\over \sqrt{2}} \pmatrix{-i\cr 1\cr 0\cr}\cr
{E^{(0)}\over \sqrt{2}} \pmatrix{1\cr +i\cr 0\cr}\cr} e^{-i\phi}\,,
\label{bd}
\end{equation}
If $B^{(0)}=E^{(0)}$ this formula describes the right-polarized radiation.
Of course, a similar formula can be written for the left-polarized
radiation. These transverse solutions can been re-written to the real
fields. For instance, Comay presented them in the following
way~[16c] in the reference frame $\Sigma$:
\begin{eqnarray}
{\bf E}_\perp &=& \cos [\omega (z-t)] \hat{\bf i}
- \sin [\omega (z-t)] \hat {\bf j}\,,\\
{\bf B}_\perp &=& \sin [\omega (z-t)] \hat{\bf i}
+ \cos [\omega (z-t)] \hat {\bf j}\,,\label{btr}
\end{eqnarray}
and analized the addition of ${\bf B}_{\vert\vert}=\sqrt{2} \hat {\bf k}$
to (\ref{btr}). Making boost to other frame of reference $\Sigma^\prime$
he claimed that a) ${\bf B}^{(3)\,\prime}$ is {\it not} parallel to the
Poynting vector; b) with the Evans postulates ${\bf E}^{(3)\,\prime}$
has a real part; c) transverse fields change, whereas ${\bf B}^{(3)}$
is left unchanged when the boost is done to the frame moving in the $z$
direction.  Comay concludes that these observations disprove the Evans
claims on these particular questions. Furthermore, he claimed that
the ${\bf B}^{(3)}$ model is {\it inconsistent} with the Relativity
Theory.

According to~\cite[Eq.(11.149)]{Jackson} the Lorentz transformation rules
for electric and magnetic fields are the following:
\begin{eqnarray}\label{l1}
{\bf E}^\prime &=& \gamma ({\bf E} + c {\mbf \beta} \times {\bf B} )
-{\gamma^2 \over \gamma+1} {\mbf\beta} ({\mbf\beta}\cdot {\bf
E})\,,\\ \label{l2} {\bf B}^\prime &=& \gamma ({\bf B} -{\mbf \beta}
\times {\bf E}/c ) -{\gamma^2 \over \gamma+1} {\mbf\beta}
({\mbf\beta}\cdot {\bf B})\,, \end{eqnarray}
where ${\mbf\beta} ={\bf
v}/c$\,,\,\,$\beta = \vert {\mbf\beta}\vert = \mbox{tanh} \phi$\,,\,
$\gamma ={1\over \sqrt{1-{\mbf\beta}^2}} =\mbox{cosh} \phi$, with
$\phi$ being the parameter of the Lorentz boost.  We shall further use the
natural unit system $c=\hbar =1$. After introducing the spin matrices
$({\bf S}^i)_{jk} =-i\epsilon^{ijk}$ and deriving relevant relations:
$$({\bf S}\cdot {\mbf \beta})_{jk} {\bf a}^k \equiv i [{\mbf \beta}
\times {\bf a}]^j\,,$$ $${\mbf \beta}^j {\mbf \beta}^k \equiv [
{\mbf \beta}^{\,2}\openone - ({\bf S}\cdot {\mbf \beta})^2
]_{jk}\,,$$ one  can rewrite Eqs. (\ref{l1},\ref{l2}) to the form
\begin{eqnarray}
\label{l3}
{\bf E}^{i\,\prime} &=& \left ( \gamma \openone +
{\gamma^2 \over \gamma +1} \left [ ({\bf S}\cdot {\mbf \beta})^2
-{\mbf\beta}^{\,2} \right ] \right )_{ij} {\bf E}^j -i\gamma ({\bf
S}\cdot {\mbf\beta})_{ij} {\bf B}^j\,,\\ \label{l4} {\bf B}^{i\,\prime}
&=& \left ( \gamma \openone + {\gamma^2 \over \gamma +1} \left [ ({\bf
S}\cdot {\mbf \beta})^2 -{\mbf\beta}^{\,2} \right ] \right )_{ij} {\bf
B}^j +i\gamma ({\bf S}\cdot {\mbf\beta})_{ij} {\bf E}^j\,.
\end{eqnarray}
Pure Lorentz  transformations (without inversions) do
not change  signs of the phase of the field functions, so we should
consider separately properties of the set of ${\bf B}^{(1)}$ and ${\bf
E}^{(1)}$, which can be regarded as the negative-energy solutions in QFT
and of another set of ${\bf B}^{(2)}$ and ${\bf E}^{(2)}$, the
positive-energy solutions.  Thus, in this framework one can deduce from
Eqs. (\ref{l3},\ref{l4})
\begin{eqnarray} \label{l11} {\bf
B}^{i\,(1)\,\prime} &=& \left ( 1+ {\gamma^2 \over \gamma +1} ({\bf
S}\cdot {\mbf \beta})^2 \right )_{ij} {\bf B}^{j\,(1)} +i\gamma ({\bf
S}\cdot {\mbf\beta})_{ij} {\bf E}^{j\,(1)}\,,\\ \label{l21} {\bf
B}^{i\,(2)\,\prime} &=& \left ( 1 +{\gamma^2 \over \gamma +1} ({\bf
S}\cdot {\mbf \beta})^2 \right )_{ij} {\bf B}^{j\,(2)} +i\gamma ({\bf
S}\cdot {\mbf\beta})_{ij} {\bf E}^{j\,(2)} \,,\\ \label{l31} {\bf
E}^{i\,(1)\,\prime} &=& \left ( 1+ {\gamma^2 \over \gamma +1} ({\bf
S}\cdot {\mbf \beta})^2 \right )_{ij} {\bf E}^{j\,(1)} -i\gamma ({\bf
S}\cdot {\mbf\beta})_{ij} {\bf B}^{j\,(1)} \,,\\ \label{l41} {\bf
E}^{i\,(2)\,\prime} &=& \left ( 1 +{\gamma^2 \over \gamma +1} ({\bf
S}\cdot {\mbf \beta})^2 \right )_{ij} {\bf E}^{j\,(2)} -i\gamma ({\bf
S}\cdot {\mbf\beta})_{ij} {\bf B}^{j\,(2)} \,,
\end{eqnarray}
and
\begin{eqnarray} \label{l111} {\bf B}^{i\,(1)\,\prime} &=& \left (
1+\gamma ({\bf S}\cdot {\mbf\beta}) +{\gamma^2 \over \gamma +1} ({\bf
S}\cdot {\mbf \beta})^2 \right )_{ij} {\bf B}^{j\,(1)} \,,\\
\label{l211} {\bf B}^{i\,(2)\,\prime} &=& \left ( 1 -\gamma ({\bf
S}\cdot {\mbf\beta}) +{\gamma^2 \over \gamma +1} ({\bf S}\cdot {\mbf
\beta})^2 \right )_{ij} {\bf B}^{j\,(2)} \,,\\
\label{l311} {\bf
E}^{i\,(1)\,\prime} &=& \left ( 1+ \gamma ({\bf S}\cdot {\mbf\beta})
+{\gamma^2 \over \gamma +1} ({\bf S}\cdot {\mbf \beta})^2 \right )_{ij}
{\bf E}^{j\,(1)} \,,\\
\label{l411} {\bf E}^{i\,(2)\,\prime} &=& \left
( 1- \gamma ({\bf S}\cdot {\mbf\beta}) +{\gamma^2 \over \gamma +1} ({\bf
S}\cdot {\mbf\beta})^2 \right )_{ij} {\bf E}^{j\,(2)} \,,
\end{eqnarray}
(when the definitions (7) are used).
To find the transformed 3-vector ${\bf B}^{(3)\,\prime}$ is
just an algebraic exercise.  Here it is
\begin{equation} {\bf
B}^{(1)\,\prime} \times {\bf B}^{(2)\,\prime} = {\bf E}^{(1)\,\prime}
\times {\bf E}^{(2)\,\prime} = i\gamma (B^{(0)})^2 (1- {\mbf \beta}
\cdot \hat{\bf k}) \left [ \hat {\bf k} -\gamma {\mbf \beta} +{\gamma^2
({\mbf \beta}\cdot \hat {\bf k}) {\mbf \beta}\over \gamma+1} \right
]\,.\label{ltt} \end{equation}
We know  that the longitudinal mode in the
Evans-Vigier theory can be considered as obtained from {\it Definition 3}.
Thus, considering that $B^{(0)}$ transforms as zero-component of a
four-vector and ${\bf B}^{(3)}$ as space components of a
four-vector:~\cite[Eq.(11.19)]{Jackson}
\begin{eqnarray}\label{lt1}
B^{(0)\,\prime} &=& \gamma (B^{(0)} -{\mbf\beta}\cdot {\bf B}^{(3)}
)\,,\\
\label{lt2} {\bf B}^{(3)\,\prime} &=& {\bf B}^{(3)} + {\gamma -1 \over
\beta^2} ({\mbf \beta} \cdot {\bf B}^{(3)}) {\mbf\beta} - \gamma
{\mbf\beta} B^{(0)} \,,
\end{eqnarray}
we find  from
(\ref{ltt}) that the relation between transverse and longitudinal modes
preserves its form:
\begin{equation}
{\bf B}^{(1)\,\prime} \times {\bf
B}^{(2)\,\prime} = i B^{(0)\,\prime} {\bf B}^{(3)\, \ast \, \prime}\,,
\end{equation}
that may be considered as a proof of the relativistic covariance
of the ${\bf B}^{(3)}$ model.

Moreover, we used that the phase factors in the formula (\ref{bd}) are
fixed between the vector and axial-vector parts of the antisymemtric
tensor field for both positive- and negative- frequency solutions if one
wants to have pure real fields. Namely,
${\bf B}^{(1)} = +i {\bf E}^{(1)}$ and
${\bf B}^{(2)} = -i {\bf E}^{(2)}$.
As we have just seen the ${\bf B}^{(3)}$ field in this
case may be regarded as a part of a 4-vector with respect to
the pure Lorentz  transformations. We are now going to take off the
abovementioned requirement and to consider the general case:
\begin{eqnarray}
\pmatrix{{\bf B}_\perp\cr {\bf
E}_\perp\cr}^\prime &=& \Lambda \left \{ \pmatrix{\tilde{\bf B}^{(1)}\cr
\tilde{\bf E}^{(1)}\cr} e^{+i\phi} + \pmatrix{\tilde{\bf B}^{(2)}\cr
\tilde{\bf E}^{(2)}\cr} e^{-i\phi} \right \} = \nonumber\\
&&\label{ltg}\\
&=&\Lambda \left \{ \pmatrix{\tilde{\bf B}^{(1)}\cr e^{i\alpha (x^\mu)}
\tilde{\bf B}^{(1)}\cr} e^{+i\phi} + \pmatrix{\tilde{\bf B}^{(2)}\cr
-e^{i\beta (x^\mu)}\tilde{\bf B}^{(2)}\cr} e^{-i\phi} \right \}\, .
\nonumber
\end{eqnarray}
Our formula (\ref{ltg}) can be re-written to the formulas
generalizing (6a) and (6b) of ref.~[2] (see also above
(\ref{l111},\ref{l211})):
\begin{eqnarray}
\label{gen1}
{\bf B}^{i\,(1)\,\prime} &=&
\left ( 1+ie^{i\alpha}\gamma ({\bf S}\cdot {\mbf\beta}) +{\gamma^2
\over \gamma +1} ({\bf S}\cdot {\mbf \beta})^2 \right )_{ij} {\bf
B}^{j\,(1)} \,,\\
\label{gen2} {\bf B}^{i\,(2)\,\prime} &=& \left ( 1
-ie^{i\beta}\gamma ({\bf S}\cdot {\mbf\beta}) +{\gamma^2 \over \gamma
+1} ({\bf S}\cdot {\mbf \beta})^2 \right )_{ij} {\bf B}^{j\,(2)}
\,.
\end{eqnarray}
One  can then repeat the procedure of ref.~[2] (see the short presentation
above) and find out that the ${\bf B}^{(3)}$ field may have {\it various}
transformation laws when the transverse fields transform with the
matrix $\Lambda$ which can be extracted from (\ref{l3},\ref{l4}).  Since
the Evans-Vigier field is  defined by the formula (\ref{bc}) we
again search the transformation law for the cross product of the
transverse modes $\left [{\bf B}^{(1)} \times {\bf B}^{(2)}\right ]^\prime
=?$ with taking into account (\ref{gen1},\ref{gen2}).
\begin{eqnarray}
&&\left [ {\bf B}^{(1)} \times {\bf B}^{(2)} \right ]^{i\,\prime}  =
\\ &=&i \gamma B^{(0)} \left \{ \left [ 1- {e^{i\alpha} +
e^{i\beta} \over 2} (i{\mbf\beta} \cdot \hat{\bf k}) \right ] (1 +
{\gamma^2 ({\mbf \beta}^2 - ({\bf S}\cdot {\mbf\beta})^2 )\over
\gamma+1} )_{ij} {\bf B}^{j\,(3)} +\right.\nonumber\\ &+& \left.
i{e^{i\alpha} - e^{i\beta} \over 2} ({\bf S}\cdot {\mbf\beta})_{ij}
{\bf B}^{j\,(3)} - \gamma B^{(0)}\left [ i{e^{i\alpha} +e^{i\beta} \over
2} + e^{i(\alpha+\beta)} ({\mbf\beta}\cdot \hat{\bf k})\right ]_{ij}
{\mbf\beta}^j \right \}\, .\nonumber
\end{eqnarray}
We used again the
{\it Definition 3} that  ${\bf B}^{(3)} = B^{(0)} \hat{\bf k}$.

One can see that we recover the formula (8) of ref.~[2] (see
(\ref{ltt}) above) when the phase factors are equal to $\alpha=-\pi/2$,
$\beta=-\pi/2$. In the case
$\alpha= +\pi/2$ and $\beta=+\pi/2$, the sign of ${\mbf\beta}$ is changed
to the opposite one.\footnote{By the way, in all his papers Evans used the
choice of phase factors incompatible with the ${\bf B}$-Cyclic Theorem in
the sense that {\it not} all the components are entries of antisymmetric
tensor fields therein.  This is the main one but not the sole error of the
Evans papers and books.} But, we are
able to obtain the transformation law as for antisymmetric tensor field,
for instance when $\alpha=-\pi/2$, $\beta=+\pi/2$.\footnote{ In the case
$\alpha= +\pi/2$ and $\beta=-\pi/2$, the sign in the third term in
parentheses (formula (30) is changed to the opposite one.} Namely, since
under this choice of the phases
\begin{equation} {\bf B}^{(1)\,\prime}
\times {\bf B}^{(2)\,\prime} = i\gamma \left [ B^{(0)}\right ]^2 \left (
\hat{\bf k} - {\gamma {\mbf\beta} ({\mbf \beta} \cdot \hat{\bf k})\over
\gamma + 1} + (i\hat{\bf i} \beta_y -i \hat{\bf j} \beta_x)\right ) \,,
\label{ltast} \end{equation} the formula (\ref{ltast}) and the formula for
opposite choice of phases lead precisely to the transformation laws of the
antisymmetric tensor fields:  \begin{equation} \left [ {\bf
B}^{i\,(3)}\right ]^\prime = \left ( 1\pm \gamma ({\bf S}\cdot
{\mbf\beta}) +{\gamma^2 \over \gamma +1} ({\bf S}\cdot {\mbf \beta})^2
\right )_{ij} {\bf B}^{j\,(3)}\,.  \end{equation} $B^{(0)}$ is a true
scalar in such a case.

What are  reasons  that we introduced additional phase factors in
Helmoltz bivectors? In~\cite{DVAF}  a similar problem has
been considered in the $(1/2,0)\oplus (0,1/2)$ (cf. also~\cite{DVAB,DVC}).
Ahluwalia identified additional phase factor(s) with Higgs-like fields
and proposed some relations with a gravitational potential. However,
the ${\bf E}$ field under definitions ($\alpha=-\pi/2$, $\beta=+\pi/2$)
becomes to be pure imaginary. One can also propose a model with the
corresponding introduction of phase factors in such a way that ${\bf
B}_\perp$ to be pure imaginary.  Can these transverse fields be
observable?  Can the phase factors be observable? A question of
experimental possibility of detection of this class of antisymmetric
tensor fields  (in fact, of the {\it anti-hermitian modes} on using the
terminology of quantum optics) is still open.  One should still note
that several authors discussed recently unusual configurations of
electromagnetic fields~\cite{Lakhb,Barr}.

Let us now look for relations with old formalisms.
The equations (10)
of~\cite{Og} is read
\begin{equation} f_{\mu\nu} ({\bf p}) \sim [
\epsilon_\mu^{(1)} ({\bf p}) \epsilon_\nu^{(2)} ({\bf p})
- \epsilon_\nu^{(1)} ({\bf p})
\epsilon_\mu^{(2)} ({\bf p})]\,\label{nt}
\end{equation}
for antisymmetric tensor $f_{\mu\nu}$ expressed through
cross product of polarization vectors in the momentum
space. This is a generalized case comparing with
the Evans-Vigier {\it Definition 2} which is obtained
if one restricts oneself by space indices.

The dynamical equations in the Ogievetski\u{\i}-Polubarinov approach
are
\begin{equation}
\Box f_{\mu\nu} -\partial_\mu \partial^\lambda
f_{\lambda\nu} + \partial_\nu \partial^\lambda f_{\lambda\mu} =
J_{\mu\nu}\,,
\end{equation}
and the new Kalb-Ramond gauge invariance is defined
with respect to transformations
\begin{equation} \delta f_{\mu\nu} = \partial_\mu
\lambda_\nu - \partial_\nu\lambda_\mu\,.
\end{equation}
It was proven that the Ogievetski\u{\i}-Polubarinov equations are related
to the Weinberg $2(2j+1)$ formalism~\cite{W2,DVWE} and~[7b-f,i].

Furthermore, they~\cite{Og} also claimed
``In the limit $m\rightarrow 0$ (or $v\rightarrow c$) the helicity becomes
a relativistic invariant, and the concept of spin loses its meaning. The
system of $2s+1$ states is no longer irreducible; it decomposes and
describes a set of different particles with zero mass and helicities $\pm
s,\pm (s-1), \ldots \pm 1,0$ (for integer spin and if parity is
conserved; the situation is analogous for half-integer
spins)\footnote{Cf. with~\cite{Kirch}. I am grateful to an anonymous
referee of {\it Physics Essays} who suggested to look for
possible connections. However, the work~\cite{Kirch} does not cite
the previous Ogievetski\u{\i}-Polubarinov statement.}."
In fact, this hints that actually the Proca-Duffin-Kemmer $j=1$
theory has {\it two} massless limit, a) the well-known Maxwell theory and
b) the {\it notoph} theory ($h=0$).
The {\it notoph} theory has been further developed by Hayashi~\cite{H}
in the context of dilaton gravity, by Kalb and Ramond~\cite{KR} in the
string context. Hundreds (if not thousands) papers exist on
the so-called Kalb-Ramond field (which is actually the {\it notoph}),
including some speculations on its connection with Yang-Mills fields.

In~\cite{phnt} I tried to use the Ogievetski\u{\i}-Polubarinov
definitions of $f_{\mu\nu}$ (see (\ref{nt})) to construct the ``potentials"
$f_{\mu\nu}$.  We can obtained for a massive field
\begin{eqnarray} f^{\mu\nu}
({\bf p}) = {iN^2 \over m} \pmatrix{0&-p_2& p_1& 0\cr p_2 &0& m+{p_r
p_l\over p_0+m} & {p_2 p_3\over p_0 +m}\cr -p_1 & -m - {p_r p_l \over
p_0+m}& 0& -{p_1 p_3\over p_0 +m}\cr 0& -{p_2 p_3 \over p_0 +m} & {p_1 p_3
\over p_0+m}&0\cr}\, , \label{lc}
\end{eqnarray}
This tensor coincides with
the longitudinal components of the antisymmetric tensor obtained in
refs.~[9a,Eqs.(2.14,2.17)] (see also below and~[7i, Eqs.(16b,17b)])
within normalizations and different forms of the spin basis.  The
longitudinal states reduce to zero in the massless case under appropriate
choice of the normalization and only if a $j=1$ particle moves along with
the third axis $OZ$.\footnote{There is also another way of thinking:
namely, to consider ``unappropriate" normalization $N=1$ and to remove
divergent part (in $m\rightarrow 0$) by a new gauge transformation.}
Finally, it is also useful to compare Eq. (\ref{lc}) with the formula (B2)
in ref.~\cite{DVALF} in order to realize the correct procedure for taking
the massless limit.

Thus, the results (at least in a
mathematical sense) surprisingly depend on a) the normalization; b) the
choice of the frame of reference.

In the  Lagrangian approach we have
\begin{equation}
{\cal L}^{Proca} = -{1\over 4} F_{\mu\nu} F^{\mu\nu} +
{m^2 \over 2} A_\mu A^\mu \Longrightarrow
{\cal L}^{Maxwell} (m\rightarrow 0)\,,
\end{equation}
and
\begin{equation}
{\cal L}^{} = -{1\over 2} F_{\mu} F^{\mu} +
{m^2 \over 4} f_{\mu\nu} f^{\mu\nu} \Longrightarrow
{\cal L}^{Notoph}= -{1\over 2} F_\mu F^\mu (m\rightarrow 0)\,,
\label{ln}
\end{equation}
where
\begin{equation}
F^\mu = {i\over 2}\epsilon^{\mu\nu\alpha\beta}
\partial_\beta f_{\nu\alpha}
= \partial_\beta \tilde f^{\mu\beta}
\end{equation}
(if one applies the duality relations). Thus, we observe that a) it is
important to consider the parity matters (the dual tensor has
different parity properties); b)  we may look for
connections with the dual electrodynamics~\cite{Strazh}.

The above surprising conclusions induced me to start form the basic
group-theoretical postulates in order to understand the origins of
the Ogievetski\u{\i}-Polubarinov-Evans-Vigier results.
The set of Bargmann-Wigner equations, ref~\cite{BW} for
$j=1$ is written, e.g., ref.~\cite{Lurie}
\begin{eqnarray} \left [
i\gamma^\mu \partial_\mu -m \right ]_{\alpha\beta} \Psi_{\beta\gamma} (x)
&=& 0\,,\label{bw1}\\ \left [ i\gamma^\mu \partial_\mu -m \right
]_{\gamma\beta} \Psi_{\alpha\beta} (x) &=& 0\,,\label{bw2}
\end{eqnarray}
where one usually uses
\begin{equation} \Psi_{\left \{ \alpha\beta \right
\}} = m\gamma^\mu_{\alpha\delta} R_{\delta\beta} A_\mu +{1\over 2}
\sigma^{\mu\nu}_{\alpha\delta} R_{\delta\beta} F_{\mu\nu}\,, \label{si}
\end{equation}
In order to facilitate an analysis of parity properties of the
corresponding fields one should introduce also the term $\sim (\gamma^5
\sigma^{\mu\nu} R)_{\alpha\beta} \tilde f_{\mu\nu}$. In order to understand
normalization matters one should put arbitrary (dimensional, in general)
coefficients in this expansion or in definitions of the fields and
4-potentials~\cite{phnt}.  The $R$ matrix is
\begin{equation}
R=\pmatrix{i\Theta & 0\cr 0&-i\Theta\cr}\quad,\quad \Theta = -i\sigma_2 =
\pmatrix{0&-1\cr 1&0\cr}\, .  \end{equation} Matrices $\gamma^{\mu}$ are
chosen in the Weyl representation, {\it i.e.}, $\gamma^5$ is assumed to be
diagonal.  The reflection operator $R$ has the properties
\begin{eqnarray}
&& R^T = -R\,,\quad R^\dagger =R = R^{-1}\,,\\ && R^{-1} \gamma^5 R =
(\gamma^5)^T\,,\\ && R^{-1} \gamma^\mu R = -(\gamma^\mu)^T\,,\\ && R^{-1}
\sigma^{\mu\nu} R = - (\sigma^{\mu\nu})^T\,.
\end{eqnarray}
They are
necessary for the expansion (\ref{si}) to be possible in such a form,
{\it i.e.}, in order the $\gamma^{\mu} R$, $\sigma^{\mu\nu} R$ and
(if considered) $\gamma^5  \sigma^{\mu\nu} R$ to be {\it symmetrical}
matrices.

I used the expansion which is similar to (\ref{si})
\begin{equation}
\Psi_{\left \{ \alpha\beta \right \}} =
\gamma^\mu_{\alpha\delta} R_{\delta\beta} F_\mu
+\sigma^{\mu\nu}_{\alpha\delta} R_{\delta\beta} F_{\mu\nu}\,,
\label{si0}
\end{equation}
and obtained
\begin{eqnarray}
&& \partial_\alpha F^{\alpha\mu} +{m\over 2} F^\mu = 0\,,\label{p1}\\
&& 2m F^{\mu\nu} = \partial^\mu F^\nu -\partial^\nu
F^\mu\quad.\label{p2}
\end{eqnarray}
If one renormalizes $F^\mu \rightarrow 2mA^\mu$ or $F_{\mu\nu} \rightarrow
{1\over 2m} F_{\mu\nu}$
one obtains ``textbooks" Proca equations. But, {\it physical contents of
the massless limits of these equations may be different}.

Let us track origins of this conclusion in detail.
If one advocates the following definitions~\cite[p.209]{Wein}
\begin{eqnarray}
\epsilon^\mu  ({\bf 0}, +1) = - {1\over \sqrt{2}}
\pmatrix{0\cr 1\cr i \cr 0\cr}\,,\quad
\epsilon^\mu  ({\bf 0}, 0) =
\pmatrix{0\cr 0\cr 0 \cr 1\cr}\,,\quad
\epsilon^\mu  ({\bf 0}, -1) = {1\over \sqrt{2}}
\pmatrix{0\cr 1\cr -i \cr 0\cr}
\end{eqnarray}
and ($\widehat p^i = p^i /\mid {\bf p} \mid$,\, $\gamma
= E_p/m$), ref.~\cite[p.68]{Wein} or ref.~\cite[p.108]{Novozh},
\begin{eqnarray}
&&
\epsilon^\mu ({\bf p}, h) =
L^{\mu}_{\quad\nu} ({\bf p}) \epsilon^\nu ({\bf 0},h)\,, \\
&& L^0_{\quad 0} ({\bf p}) = \gamma\, ,\quad L^i_{\quad 0} ({\bf p}) =
L^0_{\quad i} ({\bf p}) = \widehat p_i \sqrt{\gamma^2 -1}\, ,\\
&&L^i_{\quad k} ({\bf p}) = \delta_{ik} +(\gamma -1) \widehat p_i \widehat
p_k  \end{eqnarray}
for the field operator of the 4-vector
potential, ref.~\cite[p.109]{Novozh} or
ref.~\cite[p.129]{Itzyk}\footnote{Remember that the invariant integral
measure over the Minkowski space for physical particles is $$\int d^4 p
\delta (p^2 -m^2)\equiv \int {d^3  {\bf p} \over 2E_p}\quad,\quad E_p =
\sqrt{{\bf p}^2 +m^2}\quad.$$ Therefore, we use the field operator as in
(\ref{fo}). The coefficient $(2\pi)^3$ can be considered at this stage as
chosen for convenience.  In ref.~\cite{Wein} the factor $1/(2E_p)$ was
absorbed in creation/annihilation operators and instead of the field
operator (\ref{fo}) the operator was used in which the $\epsilon^\mu
({\bf p}, h)$ functions for a massive spin-1 particle were
substituted by $u^\mu ({\bf p}, h) = (2E_p)^{-1/2} \epsilon^\mu ({\bf
p}, h)$, which may lead to confusions in searching
massless limits $m\rightarrow 0$ for  classical polarization
vectors.}$^,$\footnote{In
general, it may be useful to consider front-form helicities (or
``time-like" polarizations) too.  But, we leave a presentation of
a rigorous theory of this type for subsequent publications.}
\begin{equation} A^\mu (x) =
\sum_{h=0,\pm 1} \int {d^3 {\bf p} \over (2\pi)^3 }
{1\over 2E_p} \left
[\epsilon^\mu ({\bf p}, h) a ({\bf p}, h) e^{-ip\cdot x} +
(\epsilon^\mu ({\bf p}, h))^c b^\dagger ({\bf p}, h) e^{+ip\cdot
x} \right ]\,, \label{fo}
\end{equation}
the normalization of the wave
functions in the momentum representation is thus chosen to the unit,
$\epsilon_\mu^\ast ({\bf p}, h) \epsilon^\mu ({\bf p},h) = -
1$.\footnote{The metric used in this paper $g^{\mu\nu} = \mbox{diag} (1,
-1, -1, -1)$ is different from that of ref.~\cite{Wein}.} \, We observe
that in the massless limit all defined polarization vectors of the
momentum space do not have good behaviour; the functions describing spin-1
particles tend to infinity.  This is not satisfactory, in my opinion,
even though one can still claim that singularities may be  removed by
rotation and/or choice of a gauge parameter.  After renormalizing the
potentials, {\it e.~g.}, $\epsilon^\mu \rightarrow u^\mu \equiv m
\epsilon^\mu$ we come to the field functions in the momentum
representation:
\begin{equation}
u^\mu
({\bf p}, +1)= -{N\over \sqrt{2}m}\pmatrix{p_r\cr m+ {p_1 p_r \over
E_p+m}\cr im +{p_2 p_r \over E_p+m}\cr {p_3 p_r \over
E_p+m}\cr}\,,\,\,\,  u^\mu ({\bf p}, -1)= {N\over
\sqrt{2}m}\pmatrix{p_l\cr m+ {p_1 p_l \over E_p+m}\cr -im +{p_2 p_l \over
E_p+m}\cr {p_3 p_l \over E_p+m}\cr}\,,\label{vp12}
\end{equation}
\begin{equation}
\qquad \qquad \qquad u^\mu ({\bf
p}, 0) = {N\over m}\pmatrix{p_3\cr {p_1 p_3 \over E_p+m}\cr {p_2 p_3
\over E_p+m}\cr m + {p_3^2 \over E_p+m}\cr}\,,  \label{vp3}
\end{equation}
($N=m$ and $p_{r,l} = p_1 \pm ip_2$) which do not
diverge in the massless limit.  Two of the massless functions (with
 $h = \pm 1$) are equal to zero when the particle, described by this
field, is moving along the third axis ($p_1 = p_2 =0$,\, $p_3 \neq 0$).
The third one ($h = 0$) is
\begin{equation} u^\mu (p_3, 0)
\mid_{m\rightarrow 0} = \pmatrix{p_3\cr 0\cr 0\cr {p_3^2 \over E_p}\cr}
\equiv  \pmatrix{E_p \cr 0 \cr 0 \cr E_p\cr}\quad, \end{equation}
\setcounter{footnote}{0}
and at
the rest ($E_p=p_3 \rightarrow 0$) also vanishes. Thus, such a field
operator describes the ``longitudinal photons" which is in complete
accordance with the Weinberg theorem $B-A= h$
for massless particles (let us remind  that we
use the $D(1/2,1/2)$ representation). Thus, the change of the
normalization can lead to the change of physical content described by
the classical field (at least, comparing with the well-accepted one).  Of
course, in the quantum case one should somehow fix the form of commutation
relations by some physical principles.\footnote{I am {\it very} grateful
to the anonymous referee of my previous papers (``Foundation of Physics")
who suggested to fix them by requirements of the dimensionless nature of
the action (apart from the requirements of the translational and
rotational invariancies).}

If one uses the dynamical relations on the basis of the consideration
of polarization vectors one can find fields:
\begin{eqnarray}
{\bf B}^{(+)} ({\bf p},
+1) &=& -{iN\over 2\sqrt{2}m} \pmatrix{-ip_3 \cr p_3 \cr ip_r\cr} =
+ e^{-i\alpha_{-1}} {\bf B}^{(-)} ({\bf p}, -1 ) \,,   \label{bp}\\
{\bf B}^{(+)} ({\bf
p}, 0) &=& {iN \over 2m} \pmatrix{p_2 \cr -p_1 \cr 0\cr} =
- e^{-i\alpha_0} {\bf B}^{(-)} ({\bf p}, 0) \,, \label{bn}\\
{\bf B}^{(+)} ({\bf p}, -1)
&=& {iN \over 2\sqrt{2} m} \pmatrix{ip_3 \cr p_3 \cr -ip_l\cr} =
+ e^{-i\alpha_{+1}} {\bf B}^{(-)} ({\bf p}, +1)
\,,\label{bm}
\end{eqnarray}
and
\begin{eqnarray}
{\bf E}^{(+)} ({\bf p}, +1) &=&  -{iN\over 2\sqrt{2}m} \pmatrix{E_p- {p_1
p_r \over E_p +m}\cr iE_p -{p_2 p_r \over E_p+m}\cr -{p_3 p_r \over
E+m}\cr} = + e^{-i\alpha^\prime_{-1}}
{\bf E}^{(-)} ({\bf p}, -1) \,,\label{ep}\\
{\bf E}^{(+)} ({\bf p}, 0) &=&  {iN\over 2m} \pmatrix{- {p_1 p_3
\over E_p+m}\cr -{p_2 p_3 \over E_p+m}\cr E_p-{p_3^2 \over
E_p+m}\cr} = - e^{-i\alpha^\prime_0} {\bf E}^{(-)} ({\bf p}, 0)
\,,\label{en}\\
{\bf E}^{(+)} ({\bf p}, -1) &=&  {iN\over
2\sqrt{2}m} \pmatrix{E_p- {p_1 p_l \over E_p+m}\cr -iE_p -{p_2 p_l \over
E_p+m}\cr -{p_3 p_l \over E_p+m}\cr} = + e^{-i\alpha^\prime_{+1}} {\bf
E}^{(-)} ({\bf p}, +1) \,,\label{em}
\end{eqnarray}
where we denoted, as previously, a normalization factor appearing in the
definitions of the potentials (and/or in the definitions of the physical
fields through potentials) as $N$.
${\bf E} ({\bf p},0)$ and ${\bf B} ({\bf p},0)$ coincide with the
strengths obtained before by different method~[9a,28], see also (\ref{lc}).
${\bf B}^\pm ({\bf p},0_t)={\bf E}^\pm ({\bf p}, 0_t) =0$ identically.
So, we again see a third component of antisymmetric tensor fields
in the massless limit which is dependent on the normalization
and rotation of the frame of reference.

However, the claim of the {\it pure}
``longitudinal nature" of the antisymmetric tensor field and/or
``Kalb-Ramond" fields after quantization still requires further
explanations.  As  one can see in~\cite{H} for a theory with ${\cal L}=
-{1\over 8} F_\mu F^\mu$ the application of the condition $(A_{ij}^{(+)}
(x)_{,j}) \vert \Psi > =0$ (in our notation $\partial_\mu f^{\mu\nu}=0$),
see the formula (18a) therein, leads to the above conclusion. Transverse
modes are eliminated by a new ``gauge" transformations. Indeed, the
expanded lagrangian is
\begin{eqnarray}
{\cal L}^{H}&=&{1\over 4}(\partial^{\mu} f_{\nu\alpha})(\partial_{\mu}
f^{\nu\alpha})-
{1\over 2} (\partial^{\mu} f^{\nu\alpha})(\partial_{\nu}
f_{\mu\alpha})=\nonumber\\
&=&-{1 \over 4}{\cal L}^{2(2j+1)}+{1\over 2}(\partial^{\mu}
f_{\alpha\mu})(\partial_{\nu} f^{\alpha\nu})\, .
\end{eqnarray}
Thus, the Ogievetski\u{\i}-Polubarinov-Hayashi Lagrangian
is equivalent to the Weinberg's Lagrangian of the $2(2j+1)$
theory~\cite{DVOL} and [7a-e],\footnote{The formal difference in
Lagrangians does not lead to physical difference. Hayashi said
that this is due to the
possiblity of applying the Fermi method {\it mutatis mutandis}.} which
is constructed as a generalization of the Dirac Lagrangian for spin 1
(instead of bispinors it contains {\it bivectors}).  In order to consider
a massive theory (we insist on making the massless limit in the end of
calculations, for physical quantities) one should add $+{1\over 4} m^2
f_{\mu\nu} f^{\mu\nu}$ as in (\ref{ln}).

The  spin operator of the massive theory, which can be found on the basis
of the N\"other formalism, is
\begin{eqnarray}
&&{\bf J}^k = {1\over 2}
\epsilon^{ijk} J^{ij} =\label{PL2}\\
&&\hspace{-10mm}=\epsilon^{ijk} \int d^3 {\bf x}
\left [ f^{0i} (\partial_\mu f^{\mu j}) + f_\mu^{\,\,\,\,j} (\partial^0
f^{\mu i} +\partial^\mu f^{i0} +\partial^i f^{0\mu} ) \right ] = {m\over
2} \int d^3 {\bf x} \, \tilde{\bf E} \times \tilde{\bf A} \,\nonumber
\end{eqnarray}
In the above equations we applied dynamical equations as
usual.  Thus, it becomes obvious, why previous authors claimed the {\it
pure} longitudinal nature of massless antisymmetric tensor field after
quantization, and why the application of the {\it generalized Lorentz
condition} leads to equating the spin operator to zero.\footnote{It is
still interesting to note that division of total angular momentum into
orbital part and spin part is {\it not} gauge invariant.}
{\it But, one should take into account the normalization issues.}
An additional mass factor in the denominator may appear a) after
``re-normalization"
${\cal L}\rightarrow {\cal L}/m^2$ (if we want to describe long-range
forces an antisymmetric tensor field must have dimension
$[energy]^2$ in the $c=\hbar=1$ unit system, and potentials, $[energy]^1$
in order the corresponding action would be dimensionless; b) due to
appropriate change of the commutation relations for creation/annihilation
operators of the higher-spin fields (including $\sim 1/m$);  c)
due to divergent terms in ${\bf E}$, ${\bf B}$, ${\bf A}$  in
$m\rightarrow 0$ under certain choice of $N$. Thus, one can recover usual
quantum electrodynamics even if we use fields (not potentials) as
dynamical variables.

\bigskip

The conclusions are:

\begin{itemize}

\item
While first experimental verifications gave negative results,
the ${\bf B}^{(3)}$ construct is theoretically possible, if one develops
it in a {\it mathematically correct way};

\item
The ${\bf B}^{(3)}$ model is a relativistic covariant model. It is
compatible with the Relativity Theory. The ${\bf B}^{(3)}$ field may be a
part of the 4-potential vector, or (if we change connections between parts
of Helmoltz bivector) may be even a part of antisymmetric tensor field;

\item
The ${\bf B}^{(3)}$ model is based on definitions which are particular
cases of the previous considerations of Ogievetski\u{\i} and Polubarinov,
Hayashi and Kalb and Ramond;

\item
The Duffin-Kemmer-Proca theory has two massless limits that seems to be in
contradictions with the Weinberg theorem ($B-A=h$);

\item
Antisymmetric tensor fields after quantization {\it may} describe
particles of both helicity $h=0$ and $h=\pm 1$ in the massless limit.
Surprisingly, the physical content depends on the normalization issues and
on the choice of the frame of reference (in fact, on rotations).

\end{itemize}

{\bf Acknowledgments.} I am thankful to Profs. A. Chubykalo, E. Comay, L.
Crowell, G.  Hunter, Y. S.  Kim, organizers and participants of the {\it
Vigier2K}, referees and editors of various journals for valuable
discussions.  I acknowledge many internet communications of Dr.  M. Evans
(1995-96) on the concept of the ${\bf B}^{(3)}$ field, while frequently do
{\it not} agree with him in many particular questions. I acknowledge
discussions (1993-98) with Dr. D.  Ahluwalia (even though I do not accept
his methods in science).

I am grateful to Zacatecas University for a professorship.
This work has been supported in part by the Mexican Sistema
Nacional de Investigadores and the Programa de Apoyo a la Carrera
Docente.

\end{document}